\documentclass{article}

\usepackage{arxiv}

\usepackage[utf8]{inputenc} 
\usepackage[T1]{fontenc}    
\usepackage{hyperref}       
\usepackage{url}            
\usepackage{booktabs}       
\usepackage{amsfonts}       
\usepackage{nicefrac}       
\usepackage{microtype}      
\usepackage{lipsum}		
\usepackage{graphicx}
\usepackage{natbib}
\usepackage{doi}

\usepackage{stfloats}
\usepackage{multirow}

\usepackage{listings}
\usepackage{xcolor}

\definecolor{codegreen}{rgb}{0,0.6,0}
\definecolor{codegray}{rgb}{0.5,0.5,0.5}
\definecolor{codepurple}{rgb}{0.58,0,0.82}
\definecolor{backcolour}{rgb}{0.95,0.95,0.92}

\setcitestyle{numbers,square}

\title{Evaluating Quantized Large Language Models for Code Generation on Low-Resource Language Benchmarks}

\author{{Enkhbold Nyamsuren}\\
	School of Computer Science and IT\\
	University College Cork\\
	Cork, Ireland, T12 XF62 \\
	\texttt{enyamsuren@ucc.ie} \\
}

\renewcommand{\shorttitle}{Evaluating Quantized Code LLMs}

\hypersetup{
pdftitle={Evaluating Quantized Large Language Models for Code Generation on Low-Resource Language Benchmarks},
pdfsubject={Programming, Artificial Intelligence, Computer Science, Software Engineering},
pdfauthor={Enkhbold Nyamsuren},
pdfkeywords={Large Language Model, Lua, code generation, quantization, AI democratization},
}

\begin{document}
\maketitle

\begin{abstract}
Democratization of AI, which makes AI accessible and usable for everyone, is an important topic, within the broader topic of the digital divide. This issue is especially relevant to Large Language Models (LLM), which are becoming increasingly popular as AI co-pilots but suffer from a lack of accessibility due to high computational demand. In this study, we evaluate whether quantization is a viable approach toward enabling LLMs on generic consumer devices. The study assesses the performance of five quantized code LLMs in Lua code generation tasks. All code LLMs had approximately 7 billion parameters and were deployed on a generic CPU-only consumer laptop. To evaluate the impact of quantization, the models were tested at 2-, 4-, and 8-bit integer precisions and compared to non-quantized code LLMs with 1.3, 2, and 3 billion parameters. Along with tasks such as question answering, text summarization, and text generation, programming tasks are one of the popular applications of AI co-pilots. Furthermore, code generation is a high-precision task, which makes it a suitable benchmark to evaluate and compare quantized models for everyday use by individuals. Lua is chosen as a low-level resource language to avoid models' biases related to high-resource languages. The results suggest that the models quantized at the 4-bit integer precision offer the best trade-off between performance and model size. These models can be comfortably deployed on an average laptop without a dedicated GPU. The performance significantly drops at the 2-bit integer precision. The models at 8-bit integer precision require more inference time that does not effectively translate to better performance. The 4-bit models with 7 billion parameters also considerably outperform non-quantized models with lower parameter numbers despite having comparable model sizes with respect to storage and memory demand. While quantization indeed increases the accessibility of smaller LLMs with 7 billion parameters, these LLMs demonstrate overall low performance (less than 50\%) on high-precision and low-resource tasks such as Lua code generation. While accessibility is improved, usability is still not at the practical level comparable to foundational LLMs such as GPT-4o or Llama 3.1 405B.
\end{abstract}


\section{Introduction}
\label{introduction}

Since the transformers were first proposed in 2017 in the seminal study by Vaswani et al. \cite{vaswani2017attention}, there have been significant advancements in the development of Artificial Intelligence. Soon after, in less than a year, OpenAI published the first GPT (Generative Pre-trained Transformer) model that demonstrated an improvement of over 5\% over existing best solutions \cite{radford2018improving}. Along with BERT (Bidirectional Encoder Representations from Transformers) \cite{devlin2018bert}, it became one of the first Large Language Models (LLM) as we know them today. Since then transformer-based large language models experienced rapid development.

When ChatGPT, powered by OpenAI's GPT-3.5, became accessible to the public in 2022, it convincingly demonstrated its capability to assist humans in various tasks. Today, many attempts are trying to leverage the power of large (language) models in a variety of applications \cite{hadi2024large, kaddour2023challenges}. Specific examples include using LLMs as AI-tutors in education \cite{kasneci2023chatgpt}, clinical decision support systems \cite{thirunavukarasu2023large}, and coding co-pilots \cite{liu2024your}.

Because LLMs are finding increasing adoption in everyday life, it is not far-fetched to assume that access to AI co-pilots can soon dictate how productive and, therefore, successful a person or an organization is. We have already experienced how increasing dependency on technology can lead to the digital divide \cite{lythreatis2022digital}, people not being able to enjoy the same opportunities because of lack of access to ICT. For example, the earliest form of the digital divide was and still is concerning the Internet in the early 2000s. Lythreatis, Singh, and El-Kassar \cite{lythreatis2022digital} mention three levels of the digital divide regarding the Internet. The first level concerns the gap in access to the Internet. The second level concerns the digital inequality in skill and knowledge necessary for using the Internet. The final level is about the overall beneficial or adverse outcomes of using the Internet. The digital divide at all three levels still persists. Despite the increasing dependency on the Internet for essential everyday activities, people still have unequal access to the Internet \cite{lai2021revisiting}. Moreover, young people are often perceived as `digital natives', yet they are also victims of the digital divide lacking both access and skills \cite{ben2022ict}.

The three levels of the digital divide initially attributed to the Internet are also highly relevant to AI. If AI adoption follows the same trend then inequality in access to AI can lead to a wider digital divide \cite{chen2024advancing, ahmed2020democratization}. Unequal access to AI, lack of skills and knowledge to effectively and efficiently use AIs, and misuse of AI can all contribute to the widening digital divide. This highlights the importance of the democratization of AI, that is how accessible AI is to ordinary individuals. Democratization is especially relevant to language models due to the issue of high computational demand that cannot be easily resolved by personal users.

In this study, we explore AI with respect to the first level of the digital divide. That is how accessible Large Language Models for code generation are to regular users. For this purpose, we evaluate the performance of smaller LLMs with less than 10 billion parameters in code generation tasks. We have specifically chosen the Lua programming language as a testbed. Lua is an example of a low-resource language \cite{cassano2023knowledge}, a language with a relatively small size of data for training LLMs. For this reason, we can avoid the bias toward high-resource languages and obtain a more representative performance evaluation of LLMs in the code generation task across different languages.

One of the main focuses of this study is to evaluate LLMs on a typical consumer device, e.g. a laptop. The mainstream online solutions, such as GPT-4o and Claude 3.5, are often pay-walled and/or have usage restrictions. Furthermore, these are black box models that have certain privacy and security risks. On the other hand, there are free and open-source alternatives such as Llama 3.1 \cite{llama3modelcard}. However, even these open-source models can be too demanding to be deployed on consumer hardware. To make LLMs more accessible, a post-training quantization \cite{jin2024comprehensive} can be applied to them. A quantized model is a compressed model with lessened demand for computing resources at the cost of reduced performance, aka quality of generated output. Depending on the number of parameters, quantized models can be deployed and run reasonably well on consumer devices.

As discussed earlier, quantization can result in a degraded performance. However, there is a distinct lack of studies exploring the effects of quantization on the performance of LLMs for code generation, or code LLMs for short. In this study, we evaluate how well LLMs quantized at different precision levels perform on Lua-based code generation benchmarks. More specifically, we infer how the quantization level affects the correctness of generated solutions, inference time, and types of errors produced. 
 
Finally, it may be possible that non-quantized models with a smaller number of parameters may perform better than quantized models with more parameters. To verify this assumption, we compare quantized models with 7 billion parameters with half-precision models with 3 billion and fewer parameters.

Overall, this study aims to answer the following research questions:
\begin{itemize}
    \item RQ1: Which code LLMs with open source or permissible licenses can be feasibly run on consumer devices with the aid of quantization?
    \item RQ2: How does quantization precision affect code LLMs concerning the quality of generated code, inference time, and types of error in the generated code?
    \item RQ3: Which quantization precision provides a reasonable trade-off between performance degradation and decreased computational demand?
    \item RQ4: How do quantized code LLMs perform compared to non-quantized code LLMs of similar model size?
\end{itemize}

\section{Related Works}
\label{relatedworks}

\subsection{Code LLMs}
\label{codellms}

HuggingFace offers a comprehensive repository\footnote{https://huggingface.co/models} of models that are free to use. The repository has become a go-to place for anyone interested in AI models including LLMs. As such, HuggingFace is an important resource toward the democratization of AI and LLMs particularly.

More relevant to this study is the Multilingual Code Models Evaluation leaderboard\footnote{https://huggingface.co/spaces/bigcode/bigcode-models-leaderboard} hosted on the HuggingFace ecosystem. The leaderboard maintains a comprehensive list of open-source and permissively licensed code LLMs ranked according to their performance in well-established coding benchmark datasets. Based on this leaderboard, this section reviews some of the better-performing models that are not derivatives of each other. 

The well-performing code LLMs (relative to other code LLMs) include CodeLlama \cite{rozière2024codellamaopenfoundation}, CodeQwen 1.5 \cite{bai2023qwen}, DeepSeek Coder \cite{guo2024deepseekcoderlargelanguagemodel}, CodeGemma \cite{codegemmateam2024codegemmaopencodemodels}, and StarCoder2 \cite{lozhkov2024starcoder2stackv2}. These code LLMs are trained on large amounts of coding data available on the Internet with popular websites, such as Stack Overflow and GitHub, being one of the primary sources.

Many of these code LLMs are available with different parameters. For example, CodeLlama is offered with 7, 13, and 34 billion parameters. While the larger number of parameters increases the model's performance, it also increases computational demand. For example, CodeLlama 34B may require somewhere between 60-70GB of memory, while CodeLlama 7B may need about 13GB of memory.

All five models listed here are free to use and have either open-source or permissive licenses making them especially suitable for individual users. These are also multilingual code LLMs meaning that the models can generate code in more than one programming language. For example, StarCoder2 supports 17 programming languages while CodeQwen 1.5 claims to support 92 programming languages.

The programming tasks can be roughly divided into three categories \cite{chai2024mceval}. The first is code generation given a natural language prompt. The second is code completion where a snippet of code is provided as a prompt and the model is expected to fill in the missing part. Finally, code understanding is the ability of a model to understand and explain a snippet of code. This study is only concerned with code generation, and the other two categories are ignored. The ranking on the Multilingual Code Models Evaluation leaderboard is also based on the code generation tasks.

\subsection{Quantization}
\label{quant}

Quantization is a compression of a Large Language Model to reduce its size with the least impact on its performance \cite{jin2024comprehensive}. There are several benefits of applying quantization to LLMs including reduced computational demands and greater accessibility of LLMs. There are two main methods of quantization, Post-Training Quantization (PTQ) and Quantization-Aware Training (QAT). While QAT methods apply quantization during the training process, PTQ methods are applied after the training phase. Because of the computational demand for QAT methods, PTQ methods are more prevalent \cite{zhu2023survey}. This study is therefore concerned with PTQ only.

PTQ methods can compress weights only or both weights and activations in LLMs. Training of LLMs usually happens at full precision, which uses a 32-bit floating point (FP32) to represent each weight and activation. After the training, LLMs can be deployed at half-precision (FP16) to improve their efficiency without a significant reduction in performance. In a quantized model, either weights or both weights and activations are represented with integers. For example, INT4/INT8 is a notation indicating that weights and activations were quantized to 4-bit and 8-bit integers respectively. Having integer bits also makes it more feasible to run a model on a CPU. However, quantizing activations can severely affect performance. Hence, it is common to quantize weights only and leave activations at full or half precision \cite{jin2024comprehensive}. 

Zhu, Li, Liu, et al. \cite{zhu2023survey} provides a comprehensive review of PTQ methods both weights-only and weight-activation combined. Among these methods, GPTQ \cite{frantar2022optq} and AWQ \cite{lin2024awqactivationawareweightquantization} are of particular interest since they are the primary methods with which the models on HuggingFace were quantized\footnote{https://huggingface.co/docs/transformers/en/main\_classes/quantization}. Both of these methods apply weights-only quantization.

Several studies evaluate quantized LLMs. Lee, Park, Kwon, et al. \cite{lee2024comprehensive} evaluated quantized models on 13 benchmark datasets assessing among others common sense knowledge, algorithmic reasoning, language knowledge, instruction following, mathematical reasoning, and conversational skills. The quantized models included Vicuna, Gemma, Llama 2, and Llama 3.1 with parameters ranging from 2B to 405B. The models were quantized with GPTQ, AWQ, and SmoothQuant algorithms to 4- and 8-bit integers. 
Jin, Du, Huang, et al. \cite{jin2024comprehensive} used 10 benchmark datasets on mathematical reasoning, language understanding, summarization, instruction following, etc to evaluate quantized Qwen 1.5 models with 7B - 72B parameters. The models were quantized to 2-, 3-, 4-, and 8-bit integers using the GPTQ, LLM.int8(), and SpQR algorithms.
Li, Ning, Wang, et al. \cite{li2024evaluatingquantizedlargelanguage} similarly evaluated 11 quantized models on language understanding, multi-step reasoning, instruction-following, dialogue, and other tasks. Overall, 19 benchmark datasets were used to test OPT, LLaMA2, Falcon, Bloomz, Mistral, ChatGLM, Vicuna, LongChat, StableLM, Gemma, and Mamba, with parameters ranging from 125M to 180B. 

In conclusion of this section, there is a particular lack of attention toward quantized code LLMs. Furthermore, in the above-mentioned studies, even quantized general-purpose LLMs such as Qwen 1.5 and Llama 3.1 are not tested against well-known benchmarks with programming tasks. This gap is addressed in the current study.

\subsection{Evaluation benchmarks for code LLMs}
\label{benchmarks}
HumanEval \cite{chen2021evaluating} and MBPP \cite{austin2021program} are arguably the earliest benchmarks that were created for evaluating code LLMs. These are originally benchmarks for testing code generation capabilities specifically in Python language. HumanEval contains 164 hand-written programming tasks. In each task, a model is asked to generate a function that should pass several unit tests. On average, there are 7.7 unit tests per task. In the original study \cite{chen2021evaluating}, HumanEval was used to evaluate GPT-3-based Codex models (OpenAI's model that powers Github Copilot) with 12 million to 12 billion parameters. Other applications of HumanEval include evaluating GitHub Copilot, Amazon CodeWhisperer, and ChatGPT \cite{yeticstiren2023evaluating}.

The Mostly Basic Programming Problems (MBPP) benchmark consists of 974 code-generation tasks at the beginner level. Similar to HumanEval, each task in MBPP has a prompt asking to write a Python function. The generated function is also tested against the unit tests defined in the task. MBPP was originally tested on dense left-to-right decoder-only transformer language models with 244 million to 137 billion parameters \cite{austin2021program}. Since then, it has become one of the main benchmarks for evaluating all major LLMs such as CodeLlama 7B-34B and StarCoder 15.5B \cite{guo2024stop}.  Overall, HumanEval and MBPP are widely used benchmarks, and Jiang, Wang, Shen, et al. \cite{jiang2024survey} provide a comprehensive review of both benchmarks and code LLMs. 

MultiPL-E \cite{cassano2023multipl} is a benchmark that combines HumanEval and MBPP. It is a multilingual benchmark that translated the original Python tasks into 18 other programming languages. These languages include Lua, Bash, C++, C\#, D, Go, Java, JavaScript, Julia, Perl, PHP, R, Racket, Ruby, Rust, Scalia, Swift, and TypeScript. The benchmark was used to evaluate the InCoder 6.7B, CodeGen 16.1B, and Codex 12B models. A later study \cite{cassano2023knowledge} applies MultiPL-E to evaluate fine-tuned versions of the StarCoderBase 1B, StarCoderBase 15B, and CodeLlama 34B models. Currently, MultiPL-E is arguably the most popular benchmark for testing multilingual code LLMs. Along with HumanEval, it is used to maintain the Multilingual Code Models Evaluation leaderboard.

Finally, MCEVAL \cite{chai2024mceval} is another multilingual benchmark that covers 40 languages including Lua. Unlike MultiPL-E, which translated Python-based HumanEval and MBPP to other programming languages, MCEVAL contains human-annotated tasks. MCEVAL contains three categories of programming tasks: code generation, code completion, and code understanding. The code generation tasks follow the same evaluation procedure as in MultiPL-E relying on function generation prompts accompanied with unit tests. All generation tasks are divided into \textit{easy}, \textit{middle}, and \textit{hard} difficulty levels. The original study \cite{chai2024mceval} evaluated 23 models with 7B to 72B parameters. These models include GPT-3.5-Turbo, GPT4-Turbo, GPT4-o,
Qwen1.5-Chat, Llama3, Phi-3-medium, Yi, CodeQwen 1.5 Chat, DeepSeek Coder Instruct, DeepSeek Coder 1.5 Instruct, CodeLlama Instruct, OCTOCODER, CodeShell, MagiCoder, WizardCoder, Codegemma, Codestral v0.1, Nxcode, OpenCodeInterpreter, etc.

Several studies used custom-developed benchmarks for specific domains. For example, Poldrack, Lu, and Begus \cite{poldrack2023ai} created a Python-based dataset with 32 data science coding problems. With this dataset, the authors tested ChatGPT and its underlying GPT-4 model. Liu, Le-Cong, Widyasari, et al. \cite{liu2024refining} compiled a dataset of 2033 programming tasks collected from LeetCode\footnote{https://leetcode.com/}, an online platform for learning programming. There are versions of Python and Java for each task. The dataset was also used to evaluate ChatGPT.

As can be seen, past works focused on evaluating full- or half-precision models without quantization applied to them. Many of the models (models with 7B or more parameters) mentioned in this section are too computationally intensive to be feasibly deployed on generic consumer devices. Models of smaller sizes, such as StarCoderBase 1B, demonstrate only marginal performance compared to the other models. Therefore, it still remains a question whether quantized code LLMs with a larger number of parameters can be leveraged. 

\subsection{Evaluation metrics}
\label{metrics}

The performance of a model in these benchmarks is usually evaluated with the $pass@k$ metric \cite{kulal2019spoc, chen2021evaluating}. Because the next token prediction in LLMs is a stochastic process, an LLM can generate different outputs for the same prompt. For a code LLM, if at least one correct solution (a solution that passed all unit tests defined in the task) is found in the top $k$ outputs then the LLM is considered to have passed the task. For example, with $pass@10$, a code LLM must produce a correct solution within its top 10 outputs. $pass@1$ is the strictest test where the code LLM is expected to produce a correct solution with its first attempt.

\subsection{Low- and high-resource languages}
\label{lowhigh}

Cassano, Gouwar, Nguyen, et al. \cite{cassano2023multipl} proposed a categorization of programming languages based on their frequency on GitHub and TIOBE rank. For example, C++, Java, JavaScript, and TypeScript are considered to be high-frequency languages. The other categories are \textit{medium}, \textit{low}, and \textit{niche}. Lua, Racket, Julia, D, and Bash fall into the niche category, each having less than 1\% representation on GitHub.

The low representation of these languages can result in a lack of training data for LLMs. For this reason, Cassano, Couwar, Lucchetti, et al. \cite{cassano2023knowledge} categorized programming languages into high- and low-resource languages. The former include languages such as Java, Python, and JavaScript. The latter include Julia, R, Racket, and Lua.

This reflects the bias in existing code LLMs toward high-resource languages. Even if quantized models perform well on high-resource languages, this performance may not generalize to low-level languages. In fact, because of the very nature of optimization, quantization may negatively affect most code generation for low-resource languages. Therefore, it is more informative to evaluate quantized models against low-resource languages rather than high-resource languages. Among the niche low-resource languages, the non-quantized models demonstrated the best performance on Lua \cite{cassano2023multipl, cassano2023knowledge, chai2024mceval}. Therefore, Lua presents a good balance between being a low-resource language and being a benchmark.

\section{Methodology}
\label{methodology}

\subsection{Run-time environment}
\label{runtime}

All models were run in a Python environment within a Windows 11 machine. Python 3.12.4 with Miniconda 24.4 was used. \textit{llama-cpp-python}\footnote{https://llama-cpp-python.readthedocs.io/en/latest/} package was used to load and run the quantized models within the Python environment. \textit{llama-cpp-python} provides a high-level Python interface to the \textit{llama.cpp} library written in C/C++. \textit{llama.cpp} was specifically designed for quantizing LLMs and working with quantized models in a GGUF format. Compared to the other solutions, such as the \textit{Transformers}\footnote{https://huggingface.co/docs/transformers} API from HuggingFace, \textit{llama-cpp-python} is more efficient and has the least impact on performance when working with quantized models. The library also supports both GPU-based and CPU-only inferences.

Concerning hardware, we used a consumer laptop Dell Latitude 5440 with Intel Core i5-1335U 1.30GHz with 12 CPU cores, 16GB DDR4 RAM, BG6 KIOXIA NVMe SSD, and no dedicated GPU. Therefore, all inference was done purely with the CPU. This device represents a generic work laptop that is typically used by different consumer segments such as businesses, academia, students, etc.

\subsection{Choice of LLMs}
\label{llmChoice}

This section addressed the research question \textit{RQ1}. The code LLMs were chosen based on licensing, comparative performance, and computational demand that can meet the limitations of the hardware specified in the preceding section. 

Table \ref{TmodelStats} summarizes the evaluated models. The following LLMs trained for code generation were selected for this study: DeepSeek Coder 6.7B Instruct \cite{guo2024deepseekcoderlargelanguagemodel}, CodeQwen 1.5 7B Chat \cite{bai2023qwen}, CodeLlama 7B Instruct \cite{rozière2024codellamaopenfoundation}, StarCoder2 7b \cite{lozhkov2024starcoder2stackv2}, and CodeGemma 7b \cite{codegemmateam2024codegemmaopencodemodels}. As of August 14th, 2024, these models were ranked among the top in the Multilingual Code Models Evaluation leaderboard. This leaderboard ranks multilingual code-generation models based on their performance on HumanEval \citep{chen2021evaluating} and MultiPL-E \cite{cassano2023multipl} benchmarks. 

To maximize the diversity of models, only original models were considered, and fine-tuned offshoots of these models were ignored. For example, Artigenz Coder DS 6.7B, while ranked high on the leaderboard, is a fine-tuned version of DeepSeek Coder 6.7B and was not included in this study. 

Only small models with 7 billion parameters or less were considered to ensure that the quantized models can be run reasonably well on consumer devices. Lastly, all these models employ a free-to-use model, albeit with certain restrictions (e.g., output from CodeLlama cannot be used to train other models).

For each of the five models, we tested 2-, 4-, and 8-bit integer weights-only quantized versions in a GPT-Generated Unified Format (GGUF)\footnote{https://github.com/ggerganov/ggml/blob/master/docs/gguf.md} format. All models were downloaded from HuggingFace's model repository. If multiple similar versions of the same model were available then the version with the highest download count was used. 2- and 8-bit quantizations are the most common quantizations at the lower and higher precision ends. 4-bit quantization is often recommended as a well-balanced trade-off between quality and size \cite{jin2024comprehensive}. We will test whether this observation also applies to code LLMs. This setup allows us to establish a correlation between quantization precision and performance thereby addressing the research questions \textit{RQ2} and \textit{RQ3}.

\begin{table*}[b]
\begin{tabular}{l l c c c} 
 \hline
 Model & URL & Quantization & RAM & Size \\
 \hline
 \multirow{3}{*}{CodeQwen 1.5 7B Chat} & \multirow{3}{*}{Qwen/CodeQwen1.5-7B-Chat-GGUF} & Q2K & 2.88GB & 2.84GB \\
 & & Q4KM & 4.48GB & 4.41GB \\
 & & Q8 & 7.01GB & 7.17GB \\
 \hline
 
 \multirow{3}{*}{Deepseek Coder 6.7B Instruct} & \multirow{3}{*}{TheBloke/deepseek-coder-6.7B-instruct-GGUF} & Q2K & 2.91GB & 2.63GB \\
 & & Q4KM & 4.07GB & 3.80GB \\
 & & Q8 & 6.85GB & 6.67GB \\
 \hline

 \multirow{3}{*}{CodeLlama 7B Instruct} & \multirow{3}{*}{TheBloke/CodeLlama-7B-Instruct-GGUF} & Q2K & 2.91GB & 2.63GB \\
 & & Q4KM & 4.07GB & 3.80GB \\
 & & Q8 & 6.84GB & 6.66GB \\
 \hline
 
 \multirow{3}{*}{CodeGemma 7b} & \multirow{3}{*}{bartowski/codegemma-7b-GGUF} & Q2K & 3.99GB & 3.24GB \\
 & & Q4KM & 5.73GB & 4.96GB \\
 & & Q8 & 9.12GB & 8.45GB \\
 \hline
 
 \multirow{3}{*}{StarCoder2 7b} & \multirow{3}{*}{bartowski/speechless-starcoder2-7b-GGUF} & Q2K & 2.68GB & 2.53GB \\
 & & Q4KM & 4.21GB & 4.09GB \\
 & & Q8 & 7.20GB & 7.10GB \\
 \hline
\end{tabular}
\caption{List of quantized models evaluated in this study. The URL column lists models' URLs on HuggingFace. Size indicates the size of a model's GGUF file.}
\label{TmodelStats}
\end{table*}

 

 
 

\subsection{Choice of benchmarks}
\label{bench}

Lua is a scripting language primarily for embedded use such as scripting in game engines and embedded devices \cite{ierusalimschy2007evolution}. Due to its niche application, Lua is considered a low-resource language \cite{cassano2023knowledge} compared to other languages such as Python, Java, and JavaScript. Low-resource languages are defined by the relatively smaller size of training data available for code LLMs. As a result, code LLMs are not as good at generating code in low-resource languages as in high-resource languages \cite{cassano2023multipl,cassano2023knowledge}. Therefore, quantized models are more likely to demonstrate performance degradation when generating code in low-resource languages such as Lua. Moreover, Lua employs unconventional programming paradigms, such as metatables, that are hard to directly generalize from other programming languages. Finally, using a low-resource language mitigates the effects of instruction fine-tuning that can be biased toward specific languages. For these reasons, Lua makes a good test case for quantized models and for answering the research question \textit{RQ2}.

Because Lua is a low-resource language, most of the Lua-based benchmarks were derived from Python-based benchmarks listed in Table \ref{TbenchStats}. MultiPL-E \cite{cassano2023multipl} is a multilingual benchmark that combines two Python-based benchmarks HumanEval \cite{chen2021evaluating} and MBPP \cite{austin2021program} and translates them into 18 other programming languages including Lua. The Lua version of HumanEval and MBPP contain 161 and 397 programming tasks respectively. These programming tasks were used to test the quantized models.

MCEVAL \cite{chai2024mceval} is another multilingual benchmark that covers 40 languages including Lua. MCEVAL has human-generated tasks. This is in contrast to HumanEval and MBPP tasks, which were created for Python and may not fully reflect the uniqueness of the languages to which these tasks were translated. MCEVAL contains 50 code generation tasks for the Lua language. 
We used the 50 code generation tasks to test the quantized models.

\begin{table}
\begin{tabular}{l c c} 
 \hline
 Benchmark & \# of Lua tasks & Task source \\
 \hline
 MultiPL-HumanEval & 161 & Python translation \\
 MultiPL-MBPP & 397 & Python translation \\
 MCEVAL & 50 & Human annotation \\
 \hline
 Total & 608 & \\
 \hline
\end{tabular}
\caption{Lua-based code generation tasks used in this study.}
\label{TbenchStats}
\end{table}

Each code generation task in the above benchmarks involves generating a standalone function given a prompt. An example of a prompt is shown in Listing \ref{LpromptEg}. The prompt includes a brief description of what the function should do, example calls to the function, and the expected returns from these calls. These are presented as Lua comments. The final line in the prompt contains Luo code with a function signature. A model is expected to generate the rest of the function.

\begin{lstlisting}[caption=The prompt from the HumanEval task: HumanEval\_8\_sum\_product, basicstyle=\small, label=LpromptEg]
-- For a given table of integers, return a  
-- table consisting of a sum and a product of 
-- all the integers in a table.
-- Empty sum should be equal to 0 and 
-- empty product should be equal to 1.
-- >>> sum_product({})
-- {0, 1}
-- >>> sum_product({1, 2, 3, 4})
-- {10, 24}
local function sum_product(numbers)
\end{lstlisting}

\subsection{Evaluation procedure}

As stated in the research question \textit{RQ2}, the code LLMs are evaluated against multiple metrics. Unit tests are used to facilitate the evaluation. Each code generation task includes several unit tests against which the model-generated function can be automatically tested. Listing \ref{LunitTests} demonstrates the unit tests for the HumanEval\_8\_sum\_product task from the HumanEval benchmark. Both HumanEval and MBPP benchmarks within MultiPL-E use the \textit{LuaUnit}\footnote{https://github.com/bluebird75/luaunit} third-party library for unit tests. On the other hand, MCEVAL uses the native \textit{assert} function for unit testing. To streamline and standardize the automated evaluation procedure, we translated the native assertions in MCEVAL to \textit{LuaUnit}-based assertions.

Table \ref{TmetricDescr} provides a summary of all metrics used to evaluate the performance of the quantized models. First, we measure \textit{pass@1}, which is the model's capacity to generate a correct function in its first attempt. If the function generated by the model passed all unit tests then it is assumed that the model generated a correct solution. If at least one test failed then the model generated an incorrect solution. 

With \textit{LuaUnit}, it is also possible to automatically differentiate between failed unit tests, runtime errors, and syntax errors. Failed unit tests and runtime errors are delivered within the same \textit{stdout} stream but with different error messages. Syntax errors are delivered within the \textit{stderr} stream. Lastly, if a function call lasted more than 30 seconds, it was considered a timeout error. Differentiating between these four types of errors can give more insight into the challenges of code generation by LLMs and the effect of quantization.

\begin{lstlisting}[caption=The unit tests from the HumanEval task: HumanEval\_8\_sum\_product, basicstyle=\small, label=LunitTests]
lu = require('luaunit')
function test_humaneval()
local candidate = sum_product
  lu.assertEquals(candidate({}),{0,1})
  lu.assertEquals(candidate({1,1,1}),{3,1})
  lu.assertEquals(candidate({100,0}),{100,0})
  lu.assertEquals(candidate({3,5,7}),{15,105})
  lu.assertEquals(candidate({10}),{10,10})
end
os.exit(lu.LuaUnit.run())
\end{lstlisting}

In addition to the error types, we also measured inference time. Inference time is the duration of time elapsed between the model receiving the prompt and the model finishing its output generation. While quantization does not aim to speed up LLMs, it can be a side effect of changing the inference process and quality. Hence, it may be insightful to investigate how inference time changes with quantization and correlates with solution accuracy.

The final evaluation metric is the lines of code (LOC) generated. LOC can represent the relative quality of code for correct solutions. For incorrect solutions, it can shed some light on the model's behaviors and the effect of quantization on it. LOC can also be correlated with inference time giving us a more granular understanding of dependencies.

\begin{table}
\begin{tabular}{p{0.25\linewidth} p{0.65\linewidth}}
 \hline
 Metric & Description \\
 \hline
 pass@1 & If a correct solution is generated at the first attempt \\
 Failed test & The generated function fails to pass at least one unit test \\
 Runtime error & An error occurs during the function runtime  \\
 Syntax error & A syntax error in the generated code \\
 Timeout & The function call fails to end within 30 seconds \\
 Inference time & Duration of time between receiving the prompt and ending the code generation \\
 Lines of Code & Lines of code generated for the task \\
 \hline
\end{tabular}
\caption{Performance metrics for evaluating the quantized models.}
\label{TmetricDescr}
\end{table}

\subsection{Model parameters}
\label{modeParams}

The first parameter of relevance to all models is \textit{temperature}. Temperature defines how predictable a model is. A lower temperature (\lstinline{t < 1}) makes a model more deterministic in its next-token prediction. A higher temperature makes the model more unpredictable while inferring the next token in a sequence. For \textit{pass@1}, a lower temperature is preferred \cite{cassano2023multipl}. The lower temperature also makes this study replicable. For these reasons, temperature is set at 0.1 for all models in this study.

\textit{Top-k} is the second parameter that makes a model more predictable. Top-k defines the greediness of the next token sampling by limiting to only the top-k tokens with the highest probability. In this study, top-k is set to 1 to enforce further the models' predictability and reproducibility of the study.

The last parameter is End-of-Sequence (EOS) tokens that tell the models when they should stop generating. By default, models can have different EOS tokens. For example, DeepSeek Coder uses `\lstinline{<|EOT|>}' as an EOS token, and CodeQwen uses `\lstinline{<|im_end|>}'. However, MultiPL-E introduces its own EOS tokens that are `\lstinline{\nlocal}',`\lstinline{\nfunction}',`\lstinline{\n--}', and `\lstinline{\n\n}'. We noticed that these additional EOS tokens make model outputs more succinct. Hence, we use these tokens in addition to the models' default EOS tokens.

\subsection{Source code and data}
\label{sourceCode}
The source code and data used in this study are available at \url{https://github.com/E-Nyamsuren/qLMM-Lua-Eval-Pipeline}. The source code includes Python pipelines for downloading and preparing the datasets, generating Lua code with the models, evaluating generated Lua code, and an R script for statistical analysis.


\section{Evaluation}
\label{results}

A total of 10944 tasks in Lua programming language were solved by the five code LLMs at 2-bit, 4-bit, and 8-bit quantization levels. The total inference time was approximately 87 hours.

\subsection{Pass@1 rates}
\label{correctness}

\begin{figure}
	\centering 
	\includegraphics[width=0.8\textwidth]{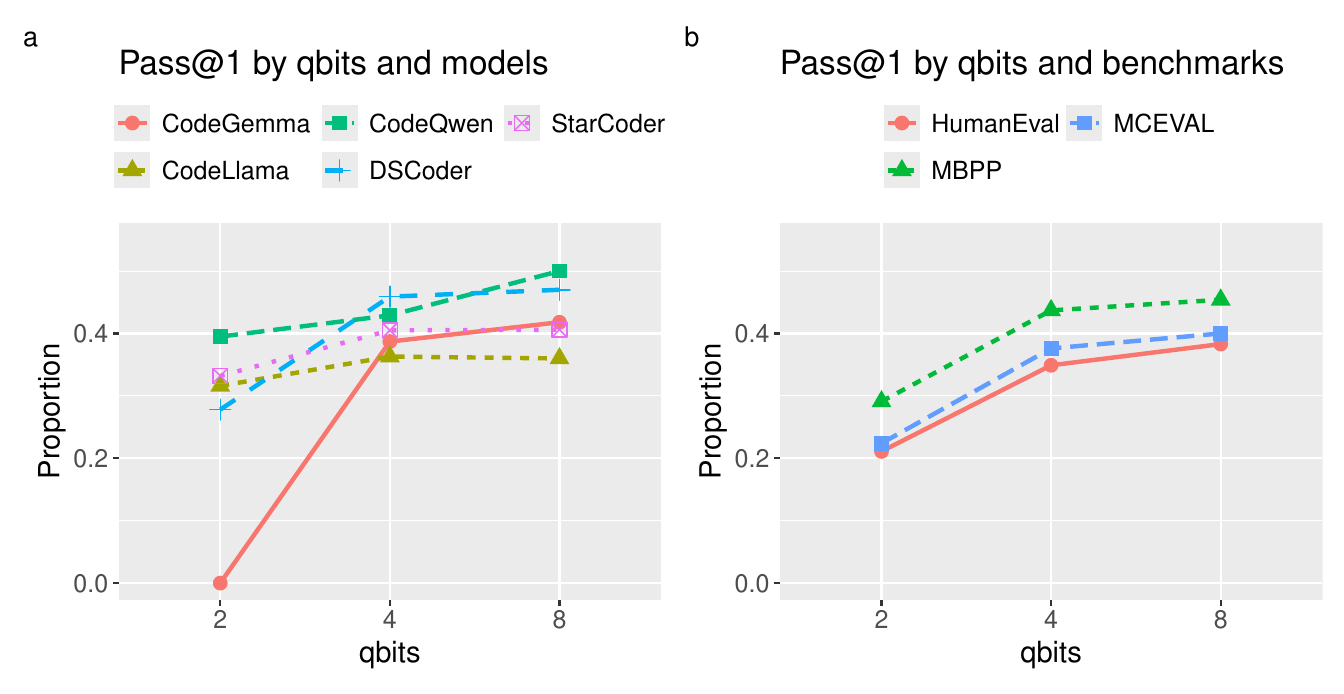}	
	\caption{Pass@1 rates (proportions of solutions that passed the unit tests) (a) of individual models and (b) by benchmarks at 2-bit, 4-bit, and 8-bit quantizations.} 
	\label{Fcorrect}%
\end{figure}

Fig. \ref{Fcorrect} depicts pass@1 rates by (a) models and quantization and by (b) benchmarks and quantization. The exact values and corresponding means are listed in Table \ref{Tcorrect}. For example, CodeLlama 7B Instruct with 2-bit integer quantization produced correct solutions for 31.6\% of tasks across all benchmarks. In the HumanEval benchmark, the correct solutions constitute 34.9\% of all solutions produced by the five models with 4-bit integer quantization. Here, a correct solution is a pass@1 solution that passed all unit tests. With the exception of 8-bit CodeQwen, the models demonstrate pass@1 rates below 50\%. It is in line with the expectation for a low-resource language like Lua.

Fig. \ref{Fcorrect}a suggests a negative effect of quantization on models' performance. The drop in performance is more evident in 2-bit models. A special case is CodeGemma, which failed to solve any task at the 2-bit quantization. The biggest performance gains are from the 2-bit model to the 4-bit models. However, compared to the 4-bit models, the 8-bit models perform only marginally better. The performance of DeepSeek Coder, StarCoder, and CodeLlama does not change much between 4-bit and 8-bit quantization. CodeQwen benefited the most from the 8-bit quantization by increasing its pass@1 rate by about 7\%. According to Fig. \ref{Fcorrect}b, in all three benchmarks, the 4-bit models perform better than the 2-bit models. This effect also holds after considering the skewed values for the 2-bit models due to CodeGemma. The performance gains level up with the 8-bit models. 

\begin{figure}
	\centering 
	\includegraphics[width=0.8\textwidth]{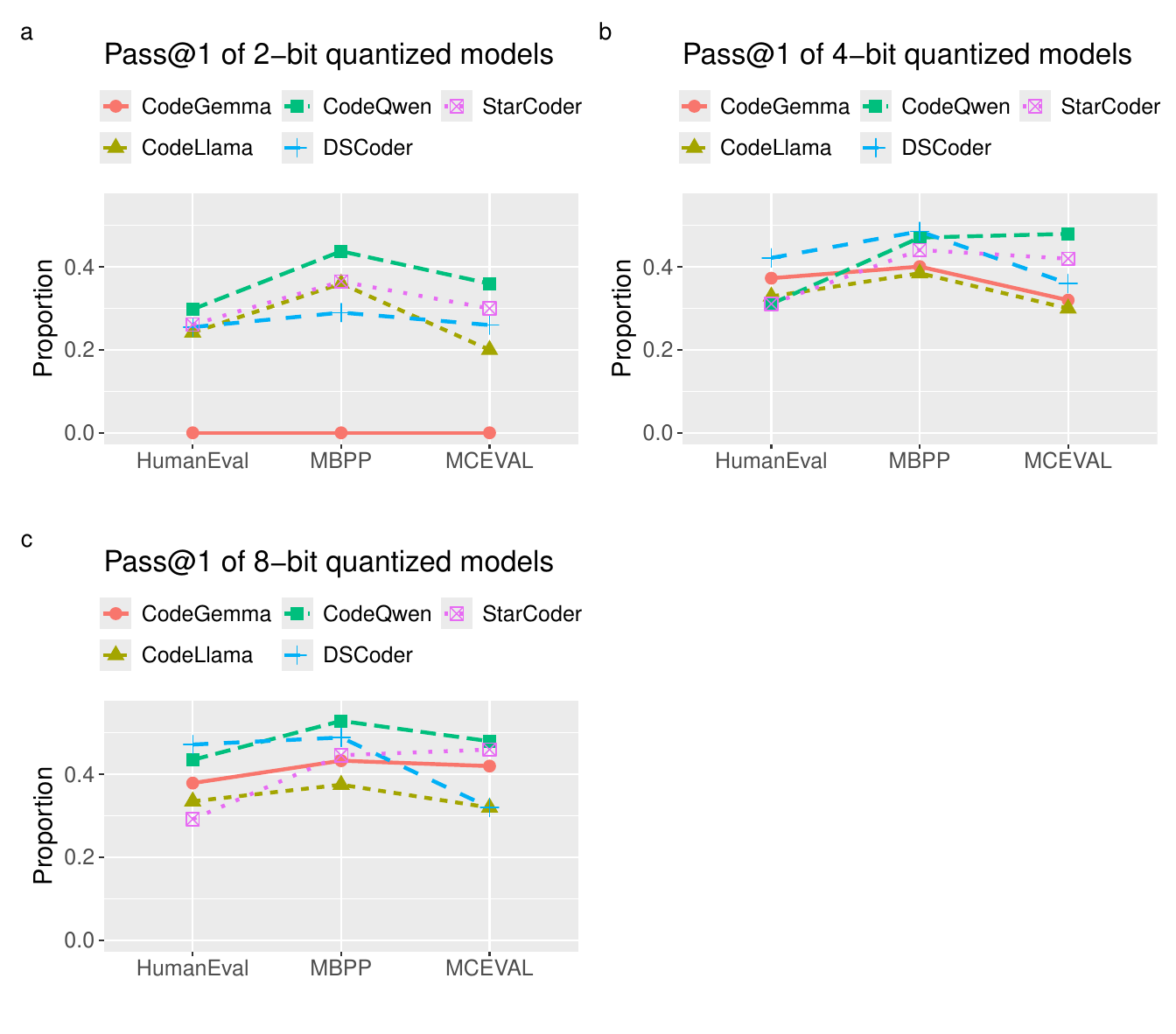}	
	\caption{Pass@1 of individual models by benchmarks at (a) 2-bit, (b) 4-bit, and (c) 8-bit quantizations.} 
	\label{FcorrByQbit}
\end{figure}

Fig. \ref{Fcorrect} also reveals that CodeQwen is consistently among the top-performing models, while CodeLlama is a poorer-performing one. True to its name, MBPP is the easiest benchmark among the three. HumanEval is slightly more difficult than MCEVAL. However, Fig. \ref{FcorrByQbit} suggests that MCEVAL may contain a greater variety of tasks highlighting the differences between the models. The pass@1 rates for the 2-bit and 4-bit models are closer to each other in HumanEval and MBPP. But for MCEVAL, the pass@1 rates remain more spread out independently of quantization. 

To statistically verify the descriptive analysis, a logistic regression was applied on pass@1. The predictors were the three benchmarks, five models, and three quantization levels. All were nominal predictors. The predictors were verified for multicollinearity with adjusted generalized VIF (GVIF). No GVIF value exceeded 2 suggesting no or low multicollinearity. Table \ref{Tpasslrm} lists the result of the regression. The intercept represents the log odds of a correct solution by 2-bit CodeGemma in the HumanEval benchmark. According to the regression, the log-odds increase for 4-bit and 8-bit quantization but the increase from 4-bit to 8-bit (0.094) is not as high as from 2-bit to 4-bit (0.668). The log odds are also significantly higher for MBPP compared to HumanEval but do not change significantly for MCEVAL. Lastly, the log odds suggest that CodeQwen and DeepSeek Coder are the overall top-performing models. CodeLlama is shown to be better than CodeGemma but it is likely due to skewed data from the 2-bit models.

The logistic regression model should be interpreted with certain care. McFadden's Pseudo $R^2$ for the regression model is 0.035 meaning that the model fails to explain a major part of variance. Also, Fig. \ref{Fcorrect} and \ref{FcorrByQbit} suggest at least two interaction effects: between quantization levels and models and between benchmarks and models. We also tested a logistic model that includes all combinations of two-way interactions between the predictors. This more complex model had only a marginal increase in McFadden's pseudo $R^2$ (0.065) and a slightly lower AIC of 11275.06 compared to the AIC of 11602.03 of the first model. Furthermore, only one interaction effect was significant in the complex model and none of the main effects was significant. This result and the overall low pseudo $R^2$ may be explained by a large variance in task difficulty that may be present in the benchmarks. This variance cannot be properly captured with the current predictors. Nevertheless, the model in Table \ref{Tpasslrm} is sufficient for exploring the effects of quantization.

\begin{table}
\begin{tabular}{p{2cm} p{2cm} c c c} 
 \hline
 \multicolumn{2}{}{} & \multicolumn{3}{c}{Quantization (qbits)} \\
 \multicolumn{2}{}{} & 2 & 4 & 8 \\
 \hline
 \multirow{6}{*}{Models} & CodeGemma & 0.000 & 0.387 & 0.418 \\
 & CodeQwen & 0.395 & 0.429 & 0.500 \\
 & StarCoder & 0.332 & 0.405 & 0.406 \\
 & CodeLlama & 0.316 & 0.363 & 0.360 \\
 & DSCoder & 0.278 & 0.459 & 0.470 \\
 & \textit{Mean} & 0.264 & 0.409 & 0.431 \\
 \hline
 \multirow{4}{*}{Benchmarks} & HumanEval & 0.211 & 0.349 & 0.383 \\
 & MBPP & 0.291 & 0.437 & 0.454 \\
 & MCEVAL & 0.224 & 0.376 & 0.400 \\
 & \textit{Mean} & 0.242 & 0.387 & 0.412 \\
 \hline
\end{tabular}
\caption{Pass@1 rates (proportions of solutions that passed the unit tests) by individual models and benchmarks at the three quantizations.}
\label{Tcorrect}
\end{table}

\begin{table}
\begin{tabular}{l r r r r} 
 \hline
 Predictor & Estimate & SE & z value & $Pr(>|z|)$ \\
 \hline
 Intercept  & -1.759 & 0.076 & -23.09 & $<0.001$ \\
 MBPP  & 0.363 & 0.052 & 6.956 & $<0.001$ \\
 MCEVAL & 0.090 & 0.091 & 0.996 & 0.319 \\
 CodeLlama & 0.380 & 0.073 & 5.191 & $<0.001$ \\
 CodeQwen & 0.791 & 0.072 & 11.008 & $<0.001$ \\
 DSCoder & 0.627 & 0.072 & 8.670 & $<0.001$\\
 StarCoder & 0.534 & 0.073 & 7.358 & $<0.001$\\
 4-bit & 0.668 & 0.056 & 11.962 & $<0.001$\\
 8-bit & 0.762 & 0.056 & 13.686 & $<0.001$\\
 \hline
\end{tabular}
\caption{The results of logistic regression on solution correctness with benchmarks, model, and q-bits as nominal predictors.}
\label{Tpasslrm}
\end{table}

\subsection{Errors}
\label{errors}

\begin{figure}
	\centering 
	\includegraphics[width=0.7\textwidth]{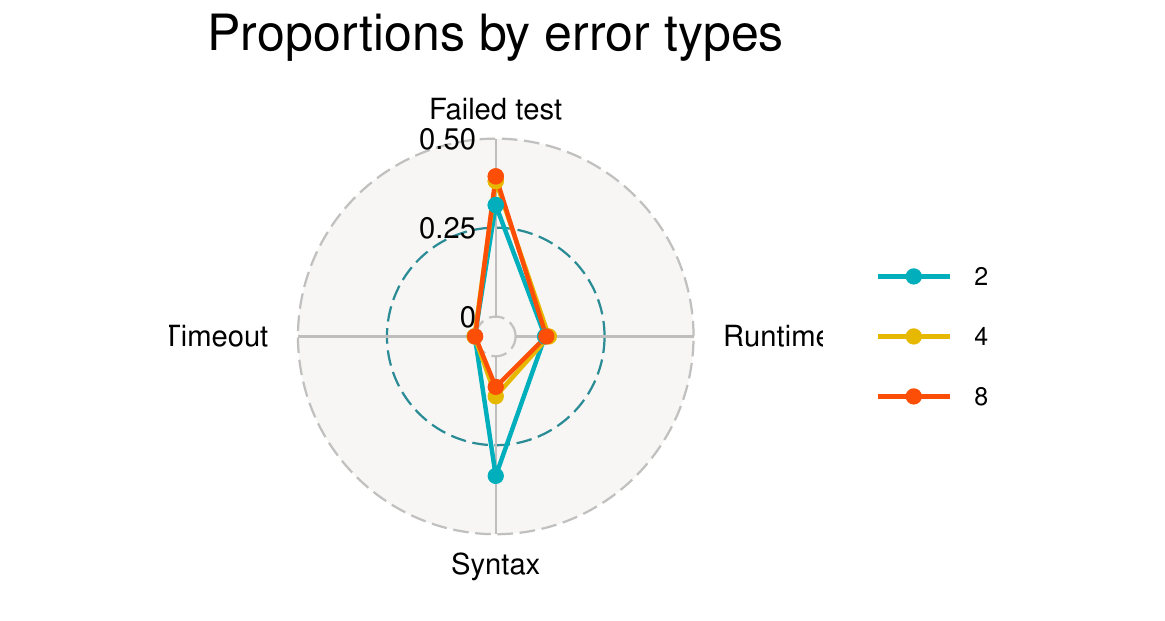}	
	\caption{Proportions of the four error types shown separately for each quantization level.} 
	\label{Ferrors}
\end{figure}

As discussed in Section \ref{methodology}, the failed solutions were categorized into failed unit tests, runtime errors, syntax errors, and timeouts. Fig. \ref{Ferrors} depicts the proportions of these categories among all solutions generated by all models in the three benchmarks and at the three quantization levels.

The failed tests constitute the majority of errors in all three quantization levels followed by the syntax errors. It should be noted that the data for 2-bit quantization is skewed because 2-bit CodeGemma was not able to solve any task due to persistent hallucinations. Listing \ref{LgemmaHall} shows example hallucinations from three tasks that demonstrate a complete breakdown of any ability to produce a coherent response.

For a more accurate view, Table \ref{Terrors} breaks down Fig. \ref{Ferrors} by the models. There is a general trend of increasing syntax errors with lower-precision quantization (lower qbits). CodeLlama seems to be less susceptible to this, but it also demonstrated poorer overall performance among the five models. DeepSeek Coder shows a decreased rate of failed tests for its 2-bit model. It is likely explained by the induction of syntax errors rather than better performance. CodeQwen is among the better-performing models. However, it seems to be more prone to syntax errors when it fails to generate a correct solution. Overall, quantization seems to affect the models' ability to produce syntax error-free code resulting in lower performance for the lower-precision models.

\begin{lstlisting}[caption=Hallucinations produced by CodeGemma at 2-bit quantizations, basicstyle=\small, label=LgemmaHall]
local function intersperse(numbers, delimeter)
 attent suspic suspic marg suspic
 suspic
 attent
 attent
 attent
 suspic marg
 ...

local function sort_numbers(numbers)
end strick strick strict

local function file_name_check(file_name)
 horrend exorbit exorbit exorbit 
 exorbit exorbit exorbit = file
\end{lstlisting}

\begin{table}[h]
\begin{tabular}{l c c c c c c} 
    \hline
    Model & Qbits & FT & RTE & SE & TO & total\\
    \hline
    CodeGemma & 2 & 0 & 0 & 1 & 0 & 1 \\
    CodeGemma & 4 & 0.37 & 0.09 & 0.15 & 0 & 0.61 \\
    CodeGemma & 8 & 0.4 & 0.07 & 0.11 & 0 & 0.58 \\
    CodeLlama & 2 & 0.52 & 0.12 & 0.04 & 0 & 0.68 \\
    CodeLlama & 4 & 0.5 & 0.09 & 0.04 & 0 & 0.63 \\
    CodeLlama & 8 & 0.51 & 0.09 & 0.04 & 0 & 0.64 \\
    CodeQwen & 2 & 0.3 & 0.07 & 0.22 & 0.01 & 0.6 \\
    CodeQwen & 4 & 0.25 & 0.05 & 0.26 & 0 & 0.56 \\
    CodeQwen & 8 & 0.28 & 0.06 & 0.16 & 0 & 0.5 \\
    DSCoder & 2 & 0.23 & 0.14 & 0.35 & 0 & 0.72 \\
    DSCoder & 4 & 0.32 & 0.14 & 0.08 & 0 & 0.54 \\
    DSCoder & 8 & 0.31 & 0.13 & 0.09 & 0 & 0.53 \\
    StarCoder & 2 & 0.52 & 0.08 & 0.07 & 0 & 0.67 \\
    StarCoder & 4 & 0.46 & 0.1 & 0.03 & 0.01 & 0.6 \\
    StarCoder & 8 & 0.47 & 0.09 & 0.03 & 0 & 0.59 \\
\hline
\end{tabular}
\caption{Proportions of failed tests (FT), runtime errors (RTE), syntax errors (SE), and timeouts (TO) among all solutions.}
\label{Terrors}
\end{table}

\subsection{Inference time}
\label{itime}

\begin{figure}
	\centering 
	\includegraphics[width=0.8\textwidth]{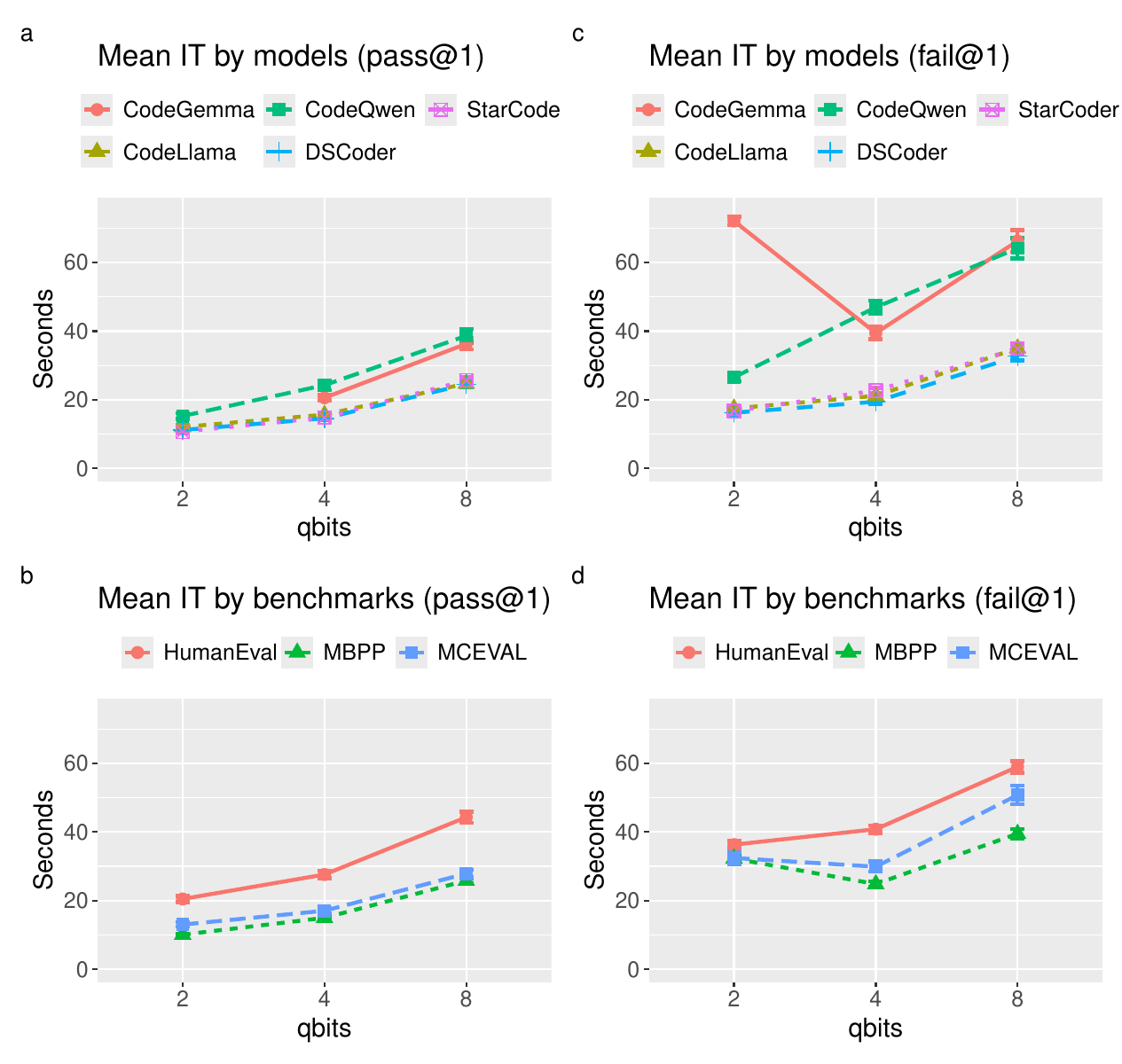}	
	\caption{Proportions of the four error types shown separately for each quantization level.} 
	\label{Fit}
\end{figure}

Fig. \ref{Fit} depicts the inference times broken down by the models and benchmarks. The figure also shows the inference times separately for the correct (pass@1) and incorrect (fail@1) solutions.

For all models, the inference time increased with higher precision (more qbits). The effect is stronger for the failed solutions. Overall, the failed solutions took longer times to generate than the correct solutions. CodeGemma, StarCoder, and DeepSeek Coder share the same pattern. CodeQwen spent more time than the other models inferring both correct and failed solutions. CodeGemma similarly demanded longer inference times but this did not as effectively translate to better results as with CodeQwen (Fig. \ref{FcorrByQbit}).

For the benchmarks, the inference times also increased with qbits. For the failed solutions in MBPP and MCEVAL, the inference times demonstrate a V-shaped pattern. It is due to CodeGemma's inflated inference times at 2-bit quantization, where it had a complete breakdown of its inference ability.

\begin{figure}
	\centering 
	\includegraphics[width=0.9\textwidth]{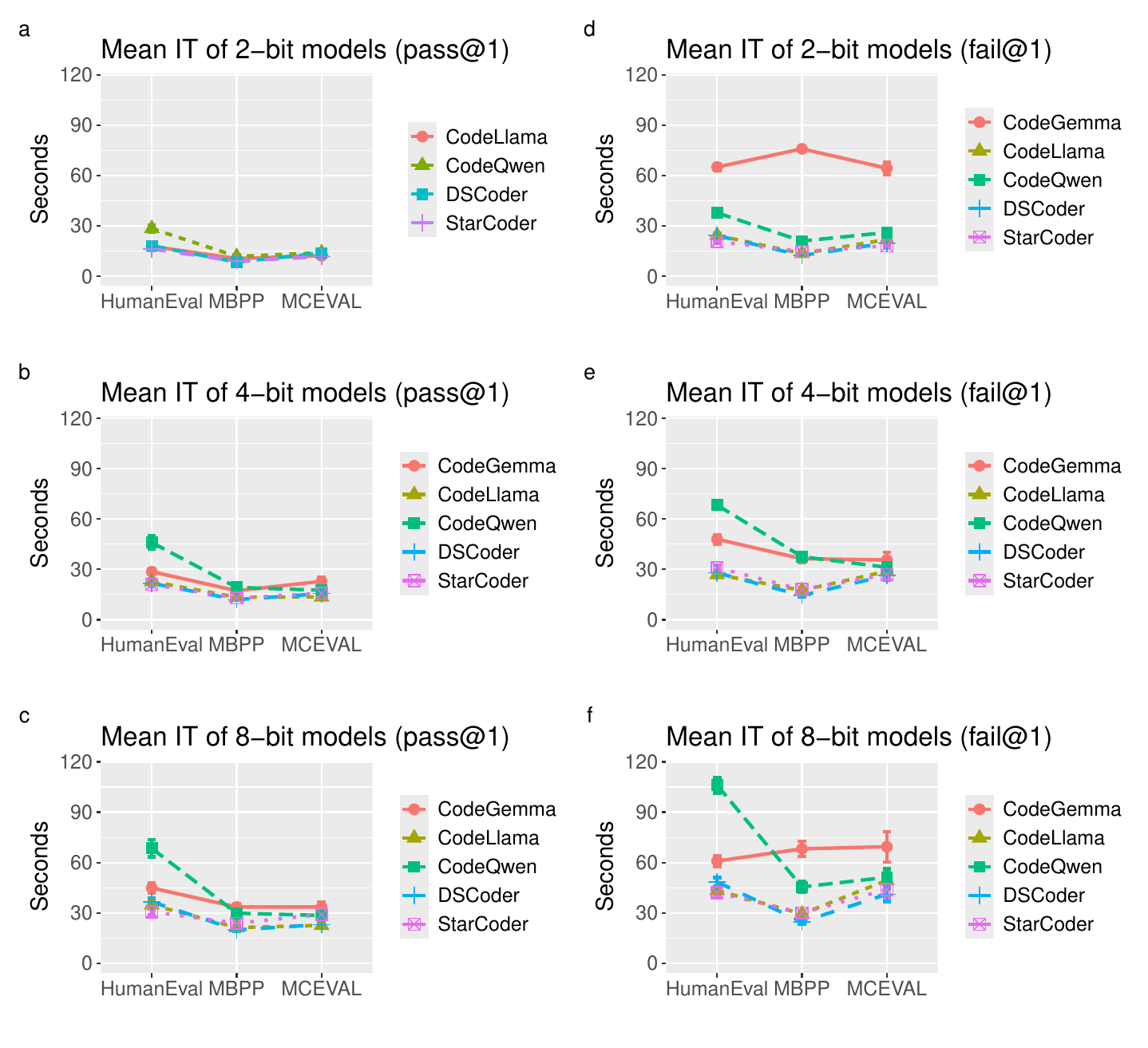}	
	\caption{Proportions of the four error types shown separately for each quantization level.} 
	\label{FitQbits}
\end{figure}

Fig. \ref{FitQbits} offers a more detailed view of the inference times. For the correct solutions, the divergence in inference times for CodeQwen from the other models is mainly observed in the HumanEval benchmark. Higher precision resulted in more divergence. According to Fig. \ref{FcorrByQbit}, 8-bit CodeQwen is the second-best-performing model on HumanEval. However, the 4-bit CodeQwen model was not able to take similar advantage of longer inference time as its performance on HumanEval was not better than that of the other models. This demonstrates a variable effect of quantization on different models and benchmarks. For example, a longer inference time does not necessarily translate to better performance and may not compensate for the lower precision of a model. This is especially evident in CodeGemma, which consistently took longer inference time but demonstrated poorer performance across all three quantization levels than the other models.

A regression analysis was done on the inference times with solution correctness (boolean), benchmarks, model, and q-bits as nominal predictors. The regression model also included all possible two-way interactions. The data from CodeGemma was excluded from the analysis to avoid the skewed results from influencing the analysis. The model's adjusted $R^2$ is 0.38 compared to the adjusted $R^2$ of 0.31 of the baseline regression model without any interactions. AIC is 63743.23 compared to the AIC of 64471.65 of the baseline model.

Table \ref{Titlrm} lists the coefficients with corresponding statistics. Insignificant two-way interactions are not listed in Table \ref{Titlrm}. The intercept represents an average inference time in seconds for an incorrect solution generated by 2-bit CodeLlama in the HumanEval benchmark. The regression model confirms the effects observed in the descriptive statistics. The correct solutions required less time to generate than the incorrect solutions. For both the correct and incorrect solutions, the inference time increases with higher q-bits. The tasks in MBPP generally require less inference time than the tasks in the two other benchmarks. This discrepancy increases with higher quantization bits. Many of the interactions account for the effects related to CodeQwen: the rate of increase in inference time with more q-bits is higher than for the other models, and this effect is even more inflated in the HumanEval benchmark.

\begin{table}
\begin{tabular}{l r r r r} 
 \hline
 Predictor & Est. & SE & t value & $Pr(>|t|)$ \\
 \hline
 Intercept & 20.61 & 1.24 & 16.56 & $<0.001$ \\
 pass & -6.15 & 1.42 & -4.32 & $<0.001$ \\
 MBPP & -6.38 & 1.29 & -4.93 & $<0.001$ \\
 MCEVAL & 3.49 & 2.22 & 1.58 & 0.115 \\
 4-bit & 8.28 & 1.46 & 5.69 & $<0.001$ \\
 8-bit & 27.27 & 1.46 & 18.66 & $<0.001$ \\
 CodeQwen & 27.14 & 1.56 & 17.36 & $<0.001$ \\
 DSCoder & 1.53 & 1.56 & 0.98 & 0.325 \\   
 StarCoder & -1.92 & 1.55 & -1.23 & 0.217 \\
 4-bit:CodeQwen & 12.16 & 1.55 & 7.85 & $<0.001$ \\
 8-bit:CodeQwen & 15.85 & 1.55 & 10.20 & $<0.001$ \\
 MBPP:4-bit & -4.89 & 1.27 & -3.86 & $<0.001$ \\
 MCEVAL:4-bit & -4.79 & 2.19 & -2.17 & 0.029 \\
 MBPP:8-bit & -13.22 & 1.27 & -10.41 & $<0.001$ \\
 MCEVAL:8-bit & -8.31 & 2.19 & -3.79 & $<0.001$ \\
 pass:4-bit & -2.94 & 1.15 & -2.56 & 0.010 \\
 pass:8-bit & -6.07 & 1.14 & -5.31 & $<0.001$ \\
 pass:CodeQwen & -10.30 & 1.31 & -7.87 & $<0.001$ \\
 pass:MBPP & 4.37 & 1.09 & 4.02 & $<0.001$ \\
 MBPP:CodeQwen & -21.49 & 1.46 & -14.71 & $<0.001$ \\
 MCEVAL:CodeQwen & -29.50 & 2.53 & -11.64 & $<0.001$ \\
 MBPP:DSCoder & -3.78 & 1.46 & -2.60 & 0.009 \\ 
 \hline
\end{tabular}
\caption{The results of the linear regression on inference time with benchmarks, model, q-bits, and correctness as nominal predictors. The regression model includes all two-way interactions but only the significant interactions are listed in the table.}
\label{Titlrm}
\end{table}

\subsection{Lines of code}
\label{loc}

\begin{figure}
	\centering 
	\includegraphics[width=0.8\textwidth]{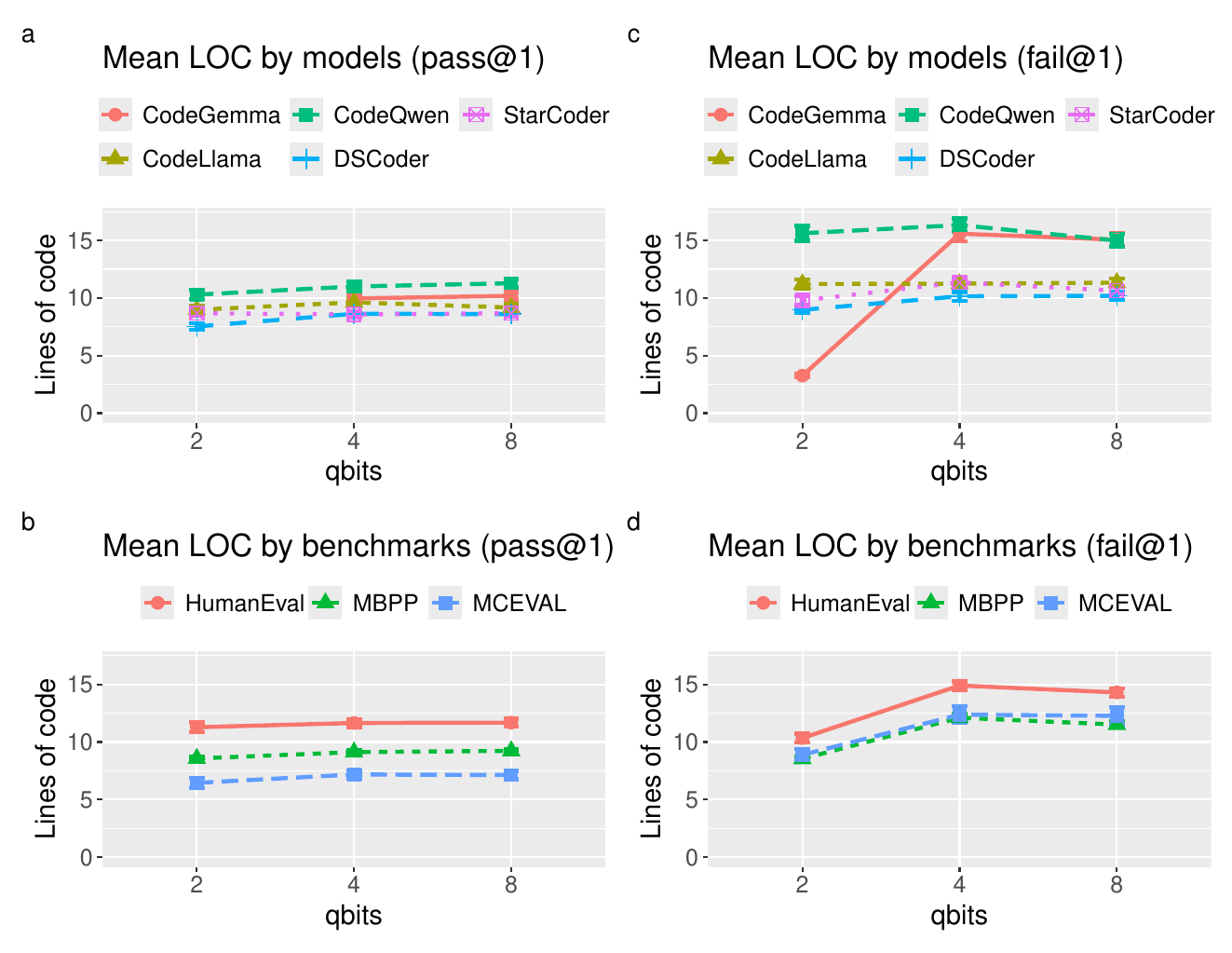}	
	\caption{Lines of code by the models and benchmarks.} 
	\label{Floc}
\end{figure}

\begin{figure*}[b]
	\centering 
	\includegraphics[width=0.98\textwidth]{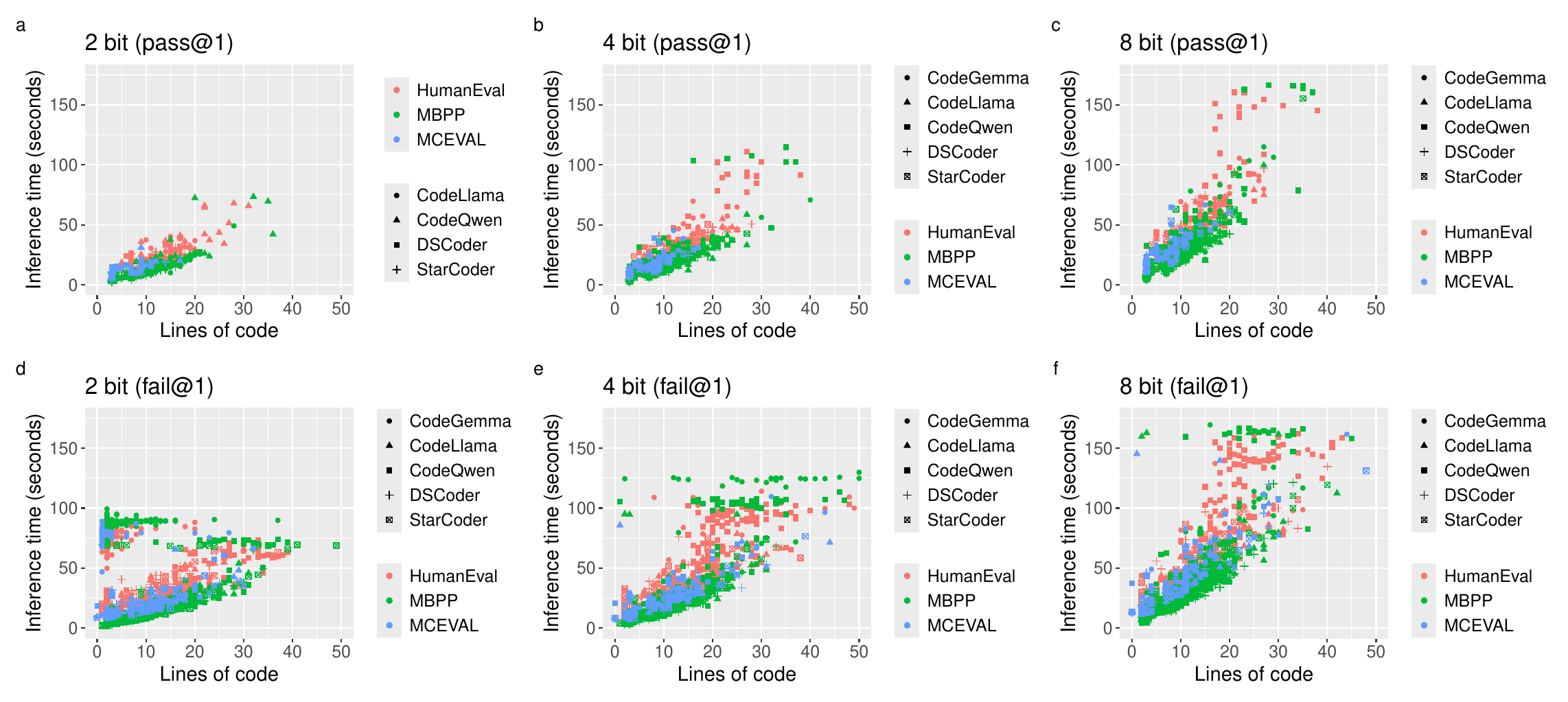}	
	\caption{Lines of Code and Inference time scatterplots.} 
	\label{FlocIt}
\end{figure*}

As shown in Fig. \ref{Floc}, lines of codes generated by the models do not differ much between the quantization levels. In generating incorrect solutions, CodeQwen and CodeGemma tended to be more verbose. The correct solutions in HumanEval require more lines of code than in the other two benchmarks. Interestingly, for the correct solutions, MBPP requires slightly more lines of code than MCEVAL while needing less inference time (Fig. \ref{Fit}). Overall, there is no effect of quantization on the number of lines of code generated. However, as depicted by Fig. \ref{FlocIt}, the time required to generate the same number of lines of code increases with higher precision quantization. This is observed for both the correct and incorrect solutions. This indicates that the increase in inference time in higher precision models is mainly due to longer forward pass time (calculations at the layers) rather than longer output generation time. In simpler terms, the higher precision models spend more time `thinking' before generating output. However, this additional thinking time does not effectively translate into better performance when the 4-bit and 8-bit models are compared (Fig. \ref{Fcorrect}).

\subsection{Comparison with FP16 models}
\label{compFp16}

Instead of using quantized models, it may be better to use a non-quantized model but with a smaller number of parameters. For this reason, we raised the research question \textit{RQ4}. We performed the same tests on DeepSeek Coder 1.3B Instruct, CodeGemma 2B, and StarCoder2 3B. The three models were tested at half-precision (FP16). The storage requirements for these models are 2.69GB, 4.40GB, and 6.06GB respectively. When loaded into memory, these models require 2.53GB, 4.44GB, and 5.79GB respectively. These sizes roughly correspond to the sizes of 2-bit, 4-bit, and 8-bit models. No low-parameter models were available for CodeLLama and CodeQwen.

\begin{figure}
	\centering 
	\includegraphics[width=0.8\textwidth]{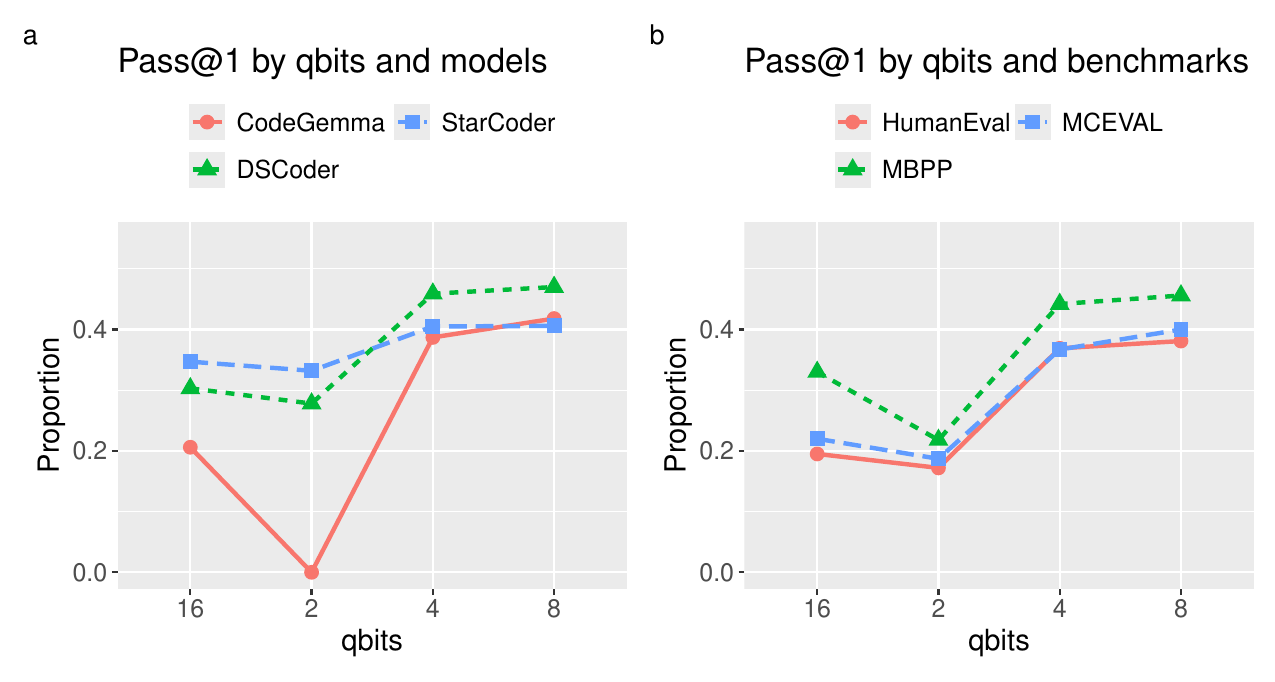}	
	\caption{Performance of the FP16 models compared to the performance of 2, 4, and 8-bit models.} 
	\label{Ffp16}
\end{figure}

As Fig. \ref{Ffp16} suggests, the low-parameter models at the FP16 half-precision performed roughly at the level of 2-bit quantized models. The low-parameter models performed considerably worse than the 4-bit quantized models.

\section{Discussion}
\label{discussion}

The results suggest that 4-bit integer quantization provides the best balance between model performance and model size. It is consistent with a conclusion made in the earlier study that evaluated quantized LLMs on general reasoning and knowledge tasks \cite{jin2024comprehensive}. Furthermore, while still being smaller in size, quantized 4-bit models with 7 billion parameters performed better than non-quantized half-precision models with 3 billion or less parameters.

On the other hand, 2-bit integer quantization resulted in a significant performance degradation. In the extreme case of 2-bit CodeGemma, there was a complete breakdown in the model's ability to generate coherent responses. This is likely an effect of hallucination \cite{yao2024llmlieshallucinationsbugs, liu2024exploringevaluatinghallucinationsllmpowered}. The low precision rounding likely impacted the model's next token prediction ability (underlying probability distributions) resulting in a sequence of repetitive out-of-context tokens.

According to \cite{cassano2023knowledge}, StarCoderBase 1B, StarCoderBase 15B, and CodeLlama 34B demonstrated MultiPL-E pass@1 percentages of 12.1, 26.6, and 43.9 for Lua. In another study \cite{cassano2023multipl}, the InCoder 6.7B, CodeGen 16.1B, and Codex 12B models demonstrated MultiPL-HumanEval pass@1 rates of approximately 0.05, 0.08, and 0.41 for Lua. In the same study, the corresponding pass@1 rates in the MultiPL-MBPP Lua benchmark were 0.13, 0.09, and 0.49. These values can be compared with the pass@1 rates in Table \ref{Tcorrect}. The 4-bit and 8-bit quantized models with 7B parameters generally do not perform much worse than the non-quantized models in \cite{cassano2023knowledge,cassano2023multipl} with higher parameter numbers. This may be explained by advances made in LLM training and fine-tuning since the studies were published. For example, we used StarCoder2 7B while \cite{cassano2023knowledge} evaluated StarCoderBase 15B. Compared to the original StarCoder released in 2023 \cite{li2023starcodersourceyou}, StarCoder2 was released in 2024 and trained on the Stack V2 dataset \cite{lozhkov2024starcoder2stackv2}, which is seven times larger than the dataset the original StarCoder was trained on.

In Chai, Liu, Yang, et al. \cite{chai2024mceval}, CodeQwen 1.5 7B Chat, DeepSeek Coder 1.5 7B Instruct, CodeLlama 7B Instruct, and Codegemma 7B it (instruct tuned) demonstrated pass@1 percentages of 48\%, 48\%, 30\%, and 48\% on the MCEVAL Lua benchmark. According to Fig. \ref{FcorrByQbit}, the 4-bit and 8-bit quantized CodeLlama Instruct models performed comparatively to the non-quantized CodeLlama Instruct model. Similarly, the 4-bit and 8-bit quantized CodeQwen 1.5 Chat only slightly underperforms the non-quantized models. There is a greater discrepancy between quantized and non-quantized CodeGemma and DeepSeek Coder models. 

This demonstrated that performance-wise the effect of quantization differs between models. Many factors may influence the quantization such as model architecture, training datasets, training process, underlying foundational models, etc. For example, the underperforming quantized models are heavily instruction fine-tuned models. The DeepSeek Coder Instruct models were pre-trained from scratch on 2 trillion tokens and further instruction-tuned on 2 billion tokens of instruction data \cite{guo2024deepseekcoderlargelanguagemodel}. CodeGemma 7B \cite{codegemmateam2024codegemmaopencodemodels} was pre-trained on 500 billion tokens but is based on the Gemma 7B model, which was pre-trained on 6 trillion tokens and further instruction-tuned on an undisclosed amount of data \cite{gemmateam2024gemmaopenmodelsbased}. In contrast, CodeQwen 1.5 is based on Qwen 1.5 pre-trained on 3 trillion tokens and was further pre-trained on 90 billion tokens \cite{bai2023qwen}. Therefore, It can be hypothesized that quantization may negatively affect most instruction fine-tuning performance. Overall, further studies are necessary to granularize the effects of quantization on different facets of code LLMs.

The study demonstrates that quantized 4-bit and 8-bit models with 7 billion parameters can be run reasonably well on laptops even without a dedicated GPU. From the perspective of computational demand, it is becoming increasingly feasible to use code LLMs for everyday programming tasks. This is especially true with the introduction of integrated environments, such as LM Studio and Ollama, that provide convenient and easy-to-use UIs for locally deploying LLMs. Generating 20 lines of correct code may require 30-50 seconds Fig. \ref{FlocIt}, which is a reasonable amount of time on a laptop oriented for business rather than coding productivity. The main problem lies with increasing inference time for generating incorrect solutions. Higher inference time does not necessarily result in better-quality code. This particularly applies to 8-bit models. Ironically, it can be another argument for using 4-bit models that can fail sooner rather than unproductively spend more time on generating incorrect solutions.

Performance-wise, both non-quantized code LLMs geared toward consumer devices and quantized code LLMs leave a lot to be desired. In most cases, these models demonstrate pass@1 rates lower than 50\% in Lua. This is very low for precision tasks such as programming. The problem is further deepened by the difficulty of detecting errors. Ironically, code LLMs are quite good at generating incorrect code that is otherwise syntactically correct and does not produce runtime errors (see Table \ref{Terrors}). Therefore, any code generated by code LLMs requires extensive testing and supervision, which may negate any advantages of using code LLMs. 

In this study, the Lua programming language was used for benchmarking the code LLMs. Lua is a low-resource language \cite{cassano2023knowledge} characterized by a lower amount of available training data compared to high-resource languages like Python and Java. Moreover, Lua has programming patterns and constructs, such as metatables and class emulations, typically not found in other languages. Hence, it is not straightforward for code LLMs to leverage generic knowledge from other languages while generating Lua code. In other words, there is no bias imposed by high-resource programming languages. Therefore, performance in Lua is more representative of the real-life performance of code LLMs on a variety of tasks. It can be further argued that real-life professional programming that code LLMs ideally need to support is about writing efficient code for specific or even niche tasks. Similar to Lua, these kinds of tasks can be seen as `low-resource tasks' even within high-resource programming languages. As such, Lua, as a niche language \cite{cassano2023multipl}, may arguably be a better representative of these `low-resource' tasks.

Likely, both proprietary and permissively licensed foundational models such as GPT-4o and Llama 3.1 405B can demonstrate significantly better performance, but accessibility is a major issue for these models. On the one hand, proprietary models like GPT-4o are pay-walled. On the other hand, permissive models like Llama 3.1 405B require a computational infrastructure that also may require considerable financial commitments. Therefore, further research is necessary to bring democratization of code LLMs. While quantization still remains a highly relevant topic, optimizing fine-tuning to be feasible on consumer devices is also essential. The ability to both fine-tune and quantize smaller LLMs for specific tasks is the necessary gap to address to enable greater consumer adoption. It should be noted that fine-tuning is not only a technical challenge. Unfortunately, datasets on which LLMs are pre-trained usually remain obscure and inaccessible to the public. Therefore, greater transparency and democratization of datasets is a necessary step toward the democratization of LLMs.

\section{Conclusions}
\label{conclusion}
Democratization of AI and Large Language Models (LLMs) in particular is becoming an increasingly relevant topic. Thanks to training on extremely large amounts of data supported by extensive computational infrastructure, LLMs can demonstrate impressive performance on a variety of tasks. However, these very same reasons became the main obstacles for individual users benefiting from LLMs. This is threatening to deepen the digital divide in the society.

Quantization is aimed at making LLMs more accessible on consumer devices by trading off performance for a lesser computational demand. In this study, we evaluated the feasibility of running code LLMs with 7 billion parameters on consumer devices. The quantized code LLMs were evaluated using Lua programming tasks as e benchmark and with respect to several metrics including pass@1 rate, inference time, error types, and lines of code generated. The quantized code LLMs were also compared to non-quantized code LLMs with lower parameter numbers.

The overall results suggest that code LLMs quantized at 4-bit integer precision can be comfortably run on an average CPU-only consumer laptop while maintaining good performance relative to other quantized and non-quantized code LLMs. The 4-bit quantized models also outperformed the non-quantized models with lower parameter numbers. 

However, the study also revealed that the exact effects of quantization are not homogeneous among the five tested models. The performance of a quantized model may also be a subject of the model architecture, pre-trained dataset, training procedure, and follow-up fine-tuning. Therefore, a more in-depth study is necessary to explore how these factors and quantization interact. Furthermore, besides the four general categories, the exact nature of errors in code generated by the code LLMs was not explored in this study. Such understanding can give a greater insight into the effects of quantization on code generation and how to mitigate respective performance degradation. This also needs to be addressed in a follow-up study.

Finally, using Lua, a low-resource language, as a benchmark further emphasizes the need to improve LLMs aimed at precision tasks such as code generation. This highlights the need for enabling consumers to not only quantize but also fine-tune models for specific tasks. Furthermore, accessibility of fine-tuning is not only an algorithmic problem but also the issue of data availability. Therefore, the democratization of LLMs also involves the democratization of training data, which also needs to be addressed in future work.

\bibliographystyle{unsrtnat}
\bibliography{manuscript}






\end{document}